\documentclass[a4paper,twocolumn,english,showpacs,nofootinbib,nobibnotes,aps,prl]{revtex4-1}
\usepackage[utf8]{inputenc}
\usepackage{amsmath}
\usepackage{amsthm}
\usepackage{amssymb}
\usepackage{graphicx}

\makeatletter

\providecommand{\tabularnewline}{\\}

\theoremstyle{plain}

  \theoremstyle{definition}
  
  \theoremstyle{plain}
  
  \theoremstyle{plain}
  
  \theoremstyle{plain}

\usepackage{braket}
\usepackage{dsfont}

\makeatother

\usepackage{babel}
  \providecommand{\corollaryname}{Corollary}
  \providecommand{\definitionname}{Definition}
  \providecommand{\lemmaname}{Lemma}
  \providecommand{\propositionname}{Proposition}
\providecommand{\theoremname}{Theorem}

\global\long\def\trace{\operatorname{Tr}}
\global\long\def\ketbra#1#2{\ket{#1}\!\bra{#2}}

\global\long\def\one{\mathds{1}}
\global\long\def\weight{\operatorname{wt}}

\newcommand{\kommentar}[1]{}

\begin{document}

\title{Constraints on correlations in multiqubit systems}

\author{Nikolai Wyderka, Felix Huber, Otfried Gühne}

\affiliation{Naturwissenschaftlich-Technische Fakult{\"a}t, 
Universit{\"a}t Siegen, Walter-Flex-Stra{\ss}e~3, 57068 Siegen, 
Germany}

\date{\today}


\begin{abstract}
The set of correlations between particles in multipartite quantum systems is 
larger than those in classical systems. Nevertheless, it is subject to 
restrictions by the underlying quantum theory. In order to better
understand the structure of this set, a possible strategy is to divide
all correlations into two components, depending on the question of whether
they involve an odd or an even number of particles.
For pure multi-qubit states we prove that these two components 
are inextricably interwoven and often one type of correlations
completely determines the other. As an application, we prove that all
pure qubit states with an odd number of qubits are uniquely determined
among all mixed states by the odd component of the correlations.
In addition, our approach leads to invariants under the time evolution
with Hamiltonians containing only odd correlations and can simplify
entanglement detection. 
\end{abstract}
\maketitle

{\it Introduction.---} 
Correlations in quantum mechanics are stronger than their counterparts in 
the classical world.  This fact is important for many applications in quantum
information processing. Taking a closer look, however, the former statement 
sounds like a truism and one realizes that many insights in quantum theory stem 
from the fact that quantum mechanical correlations are limited and not arbitrarily 
strong. For instance, the fact that in Bell experiments quantum correlations 
do not reach the values admissible by non-signaling theories has led to insightful 
discussions about underlying physical principles \cite{informationcausality, 
almostquantum}. To give another example, 
for three or more particles monogamy relations bound the entanglement between 
different pairs of particles, and their study is essential for the progress of 
entanglement theory \cite{woottersmonogamy, verstraetemonogamy, siewertmonogamy, paterek}.

If one considers multiparticle systems, however, not only correlations between
pairs of particles are relevant but also those between different sets of particles. 
Any multi-qubit state can be expressed in terms of tensor products of
Pauli matrices via the Bloch decomposition. Different terms act on different sets
of qubits, describing the correlations between just this set of particles. 
Consequently, one may ask whether there are any relations between these components
of the total correlations. For instance, for three particles, denoted by $A, B,$ 
and $C$, three different contributions can be distinguished (see also Fig.~1): 
First, there are single-body terms, acting on individual parties alone and determining
the single party density matrices. Second, there are two-body correlations acting
on the pairs $AB$, $BC$, and $CA$. Finally, there are three-body correlations 
acting on all three particles $ABC$. So the question arises: Are these three 
contributions independent of each other or is one of them determined by the 
others?

\begin{figure}[t]
\begin{centering}
 \includegraphics[width=0.85\columnwidth]{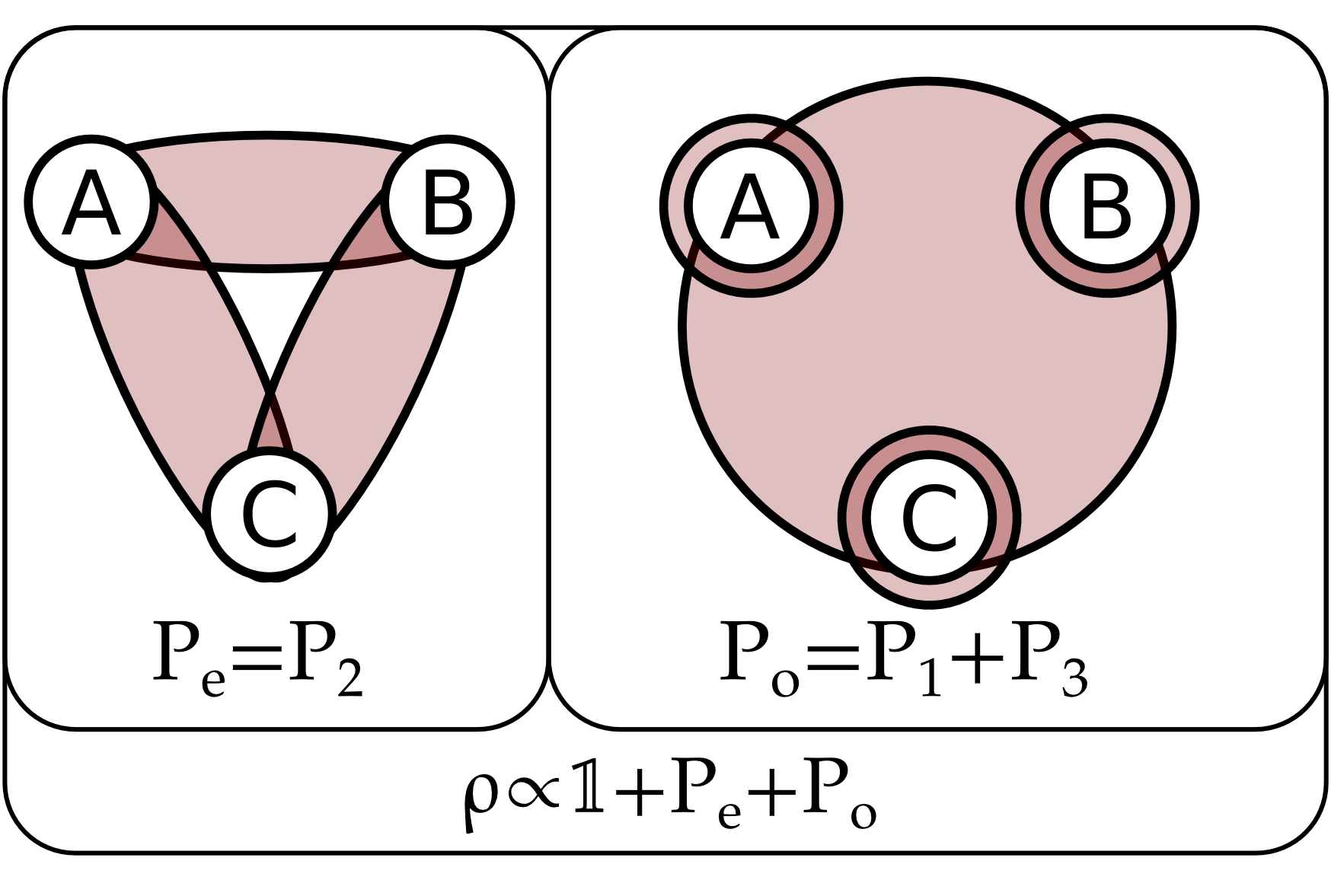}
\end{centering}
\caption{Visualization of the decomposition of a three-particle state $\rho$ 
into even and odd correlations. A state $\rho$ is expanded in Bloch 
representation as $\rho \propto \one + P_1 + P_2 + \ldots$, where $P_j$ 
denotes all terms containing $j$-body correlations. We prove that the 
even correlations $P_{\rm e}$ are determined by the odd correlations 
$P_{\rm o}$ for pure states of an odd number of qubits, so the three qubit
state is completely determined by~$P_{\rm o}$.}
\label{fig1}
\end{figure}

In this paper we present an approach to answer this and more general questions
for multi-qubit systems. We identify two components of the correlations, depending 
on the question of whether they act on an odd or even number of particles. We prove 
that the even correlations and odd correlations obey strong relations, 
one component often completely determining the
other one. Besides their fundamental interest, our results have several practical 
applications: We prove that all pure qubit states with an odd number of qubits 
are uniquely determined among all mixed states by the odd component of the 
correlations. This generalizes a famous result by Wootters and co-workers for three
particles \cite{linden2002almost}. In addition, our approach can be used to  
characterize ground states arising from Hamiltonians having even or odd interactions
only, and the time evolution 
under Hamiltonians having the odd component only. 
Finally, we apply our insights to simplify the task of
entanglement detection in certain scenarios. 

%%%%%%%%%%%%%%%%%%%%%%%%%%%%%%%%%%%%%%%%%%%%%%%%%%%%%%%%%%%%%%%%%%%%%%%
%%%%%%%%%%%%%%%%%%%%%%%%%%%%%%%%%%%%%%%%%%%%%%%%%%%%%%%%%%%%%%%%%

{\it The Bloch representation.---}
This representation is obtained by expanding an $n$-qubit quantum 
state $\rho$ in terms of tensor products of Pauli operators.
So we can write
\begin{equation}
\rho=\frac{1}{2^{n}}
\sum_{\alpha_{1},\ldots, \alpha_{n}}
c_{\alpha_{1}\ldots\alpha_{n}}
\sigma_{\alpha_{1}}\otimes\ldots\otimes\sigma_{\alpha_{n}},
\label{eq:blochdeco}
\end{equation}
where $\alpha_{i}\in\left\{ 0,1,2,3\right\}$, $\sigma_{0}=\one$,
and $\sigma_1, \sigma_2, \sigma_3$ are the usual Pauli matrices. The
coefficients $c_{\alpha_{1}\ldots\alpha_{n}}$ are given by
the expectation values
$
c_{\alpha_{1}\ldots\alpha_{n}}
=\trace(\sigma_{\alpha_{1}}\otimes\ldots\otimes\sigma_{\alpha_{n}}\rho).
$

In our approach we will sort the terms in the Bloch representation according
to the number of qubits they act on. First, we can assign to any basis element 
$\sigma_{\alpha_1} \otimes \ldots \otimes \sigma_{\alpha_n}$
its weight,
$
\weight(\sigma_{\alpha_1} \otimes \ldots \otimes \sigma_{\alpha_n}) 
:= \left|\{i\;\vert\;\alpha_i \neq 0\}\right|,
$
as the number of non-trivial Pauli matrices. Then, we group the terms 
in the decomposition according to their weight
\begin{equation}
\rho=\frac{1}{2^{n}}(\one^{\otimes n}+\sum_{j=1}^{n}P_{j})\label{eq:bloch_decomp},
\end{equation}
where $P_{j}$ denotes the sum over all contributions of weight $j$. We call 
$P_j$ also the $j$-body correlations, being determined by the expectation
values taken on groups of $j$ particles. As an example, consider the two-qubit
Bell state $\ket{\Psi^+}=(\ket{01}+\ket{10})/\sqrt{2}$, for which the corresponding 
density operator reads
$\ketbra{\Psi^+}{\Psi^+}
=
(\one\otimes\one + \sigma_{x}\otimes\sigma_{x} 
+\sigma_{y}\otimes\sigma_{y} -\sigma_{z}\otimes\sigma_{z})/4,$ 
so we have $P_{1}=0$ and 
$P_{2}=\sigma_{x}\otimes\sigma_{x}
+\sigma_{y}\otimes\sigma_{y}-\sigma_{z}\otimes\sigma_{z}$.

As our main starting point, we further group the operators according 
to the parity of their weight and define
\begin{eqnarray}
P_{\rm e} := \sum_{j \text{ even}, \; j \neq 0} P_j, \qquad P_{\rm o} 
:= \sum_{j \text{ odd}}P_j.
\end{eqnarray}
Note that $P_{0}=\one^{\otimes n}$ is excluded from $P_{\rm e}$. 
Then we can write states in the even-odd-decomposition 
(see  Fig.~\ref{fig1})
\begin{equation}
\rho = \frac1{2^n} (\one + P_{\rm e} + P_{\rm o}).
\label{evenodddecomp}
\end{equation}
The central point of our paper is that there are strong relations between
$P_{\rm e}$ and $P_{\rm o}$, and in many cases one determines the other. 
%In this way, the even and odd correlations of quantum states are fundamental
%constituents to characterize the possible correlation in quantum states.

%%%%%%%%%%%%%%%%%%%%%%%%%%%%%%%%%%%%%%%%%%%%%%%%%%%%%%%%%%%%%%%%%%%%%5
{\it State inversion.---}
Our approach is based on the state inversion map, which, for any qubit
state, can be defined as follows 
\cite{wootters1998entanglement,hill1997entanglement}:
\begin{eqnarray}
\tilde{\rho} & := \sigma_{y}^{\otimes n}\rho^{\text{T}}\sigma_{y}^{\otimes n}.
\end{eqnarray}
Physically, the state inversion is obtained by complex conjugation followed
by a spin flip. This can be represented by the anti-unitary inversion operator
$F:=(i\sigma_{y})^{\otimes n}C$ \cite{buvzek1999optimal}. Here, first the complex 
conjugation $C$ is performed and then $(i\sigma_{y})^{\otimes n}$ is applied to a pure state.
We have that $F^{\dagger}=(-1)^{n}F$ and for pure states we write
$
\ket{\tilde{\psi}} = F\ket{\psi}.
$
It follows that pure states remain pure under the state inversion. Note that 
on single-qubit Pauli matrices we have $F\sigma_{i}F^{\dagger}=-\sigma_{i}$ 
for $i\neq0$. Thus, the action of $F$ in Bloch decomposition is to flip the 
sign in front of each term that has an odd weight. Starting from Eq.~(\ref{evenodddecomp}),
we can also write
\begin{equation}
\tilde{\rho} = \frac{1}{2^{n}}(\one+P_{\rm e}-P_{\rm o}). \label{evenodddecomp2}
\end{equation}
This allows for an easier representation of the even and odd correlations, namely,
\begin{eqnarray}
 \one + P_{\rm e} = 2^{n-1}(\rho+\tilde\rho),
 \qquad P_{\rm o} = 2^{n-1}(\rho - \tilde\rho).
 \label{eqn:rhotilde_bloch}
\end{eqnarray}
The key observation is that under the state inversion, pure states of an odd 
number of qubits are mapped to orthogonal states. This fact was known before
\cite{wong2001potential, bullock2005time, osterloh2005constructing, butterley2006compatibility, designolle2017genuinely}, 
however, we give a proof that allows for generalization to qudit systems, for which
the statement is new.

\noindent
{\bf Observation 1.}
{\it For pure $n$-qudit states $\rho=\ketbra{\psi}{\psi}$ with $n$ odd we have that}
\begin{equation}
\rho\tilde\rho=0.
\label{eq:rhorhotilde0}
\end{equation}
{\it Proof.}
Let $\rho=\ketbra{\psi}{\psi}$ be the pure quantum state and denote the $n$ 
parties by $A_{1}, \dots, A_{n}$. Using the Schmidt decomposition one can 
verify that for any bipartition $M\vert \bar{M}$ of the parties one has
for the reduced state $\rho_{M}:=\trace_{\bar{M}}(\rho)$ the relation
\begin{equation}
(\rho_{M}\otimes\one_{\bar{M}})\rho=(\rho_{\bar{M}}\otimes\one_{M})\rho.\label{eq:schmidtdeco}
\end{equation}
Let us denote by $\rho_{ij\ldots} = \rho_{\{A_i,A_j,\ldots\}} \otimes \one$ 
the reduced state of parties $A_i,A_j,\ldots$, padded by identities which 
are acting trivially on the other particles. Then, state inversion can be 
written as a sum over its reductions \cite{hall2005multipartite}
\begin{equation}
\tilde{\rho}=\one-\sum_{1\leq i\leq n}\rho_{i}+\sum_{1\leq i<j\leq n}\rho_{ij}-\ldots\pm\rho. \label{state_inv_general}
\end{equation}
Note that complementary reductions have the opposite sign since $n$ 
is odd. Thus, multiplying this equation with $\rho$ and using Eq.~(\ref{eq:schmidtdeco}), every term cancels one of the others
and we have $\rho \tilde \rho = 0$.
\qed

In the qudit case, pure states do not stay pure under the state inversion,
but are mapped to positive operators. 
This generalization is studied in a later section.

\kommentar{As we will see, the operator identity in Eq.~(\ref{eq:rhorhotilde0}) has
strong implications on correlations in pure qubit states.
For even $n$, this result is not true in general.
However, there are certain states like the $W$-states for $n\geq4$,
for which the statement still holds, as will be discussed later}

%%%%%%%%%%%%%%%%%%%%%%%%%%%%%%%%%%%%%%%%%%%%%%%%%%%%%%%%%%%%%%%%%
{\it Results for an odd number of qubits.---}
Throughout this section, we consider pure qubit states of an odd number 
of parties, denoted by $\ket{\psi^\text{odd}}$.
We can directly prove our first main result.

\noindent
{\bf Observation 2.}
{\it For pure $n$-qubit states $\ket{\psi^\text{odd}}$, written in the 
even-odd decomposition as in Eq.~(\ref{evenodddecomp}), we have that

\noindent
(1) the even and odd components of the correlations commute: $[P_{\rm e},P_{\rm o}]  =  0$;}

\noindent
{\it (2) the odd correlations uniquely determine the even correlations via}
\begin{equation}
\one+P_{\rm e}=\frac{1}{2^{n-1}}P_{\rm o}^{2};\label{eq:result1}
\end{equation}
\noindent
{\it (3) the eigenvalues $\Lambda=(\lambda_1, \dots, \lambda_{2^n})$ of 
$P_{\rm e}$ and $P_{\rm o}$ are}
\begin{eqnarray}
\Lambda(P_{\rm e}) & = & (2^{n-1}-1,2^{n-1}-1,-1,\ldots,-1), \nonumber \\
\Lambda(P_{\rm o}) & = & (2^{n-1},-2^{n-1},0,\ldots,0).
\end{eqnarray}

{\it Proof.} 
We use Eq.~(\ref{eqn:rhotilde_bloch}) to write
\begin{eqnarray}
 P_{\rm o}      & = 2^{n-1}(\ketbra{\psi^\text{odd}}{\psi^\text{odd}} 
 - \ketbra{\tilde{\psi}^\text{odd}}{\tilde{\psi}^\text{odd}}), \nonumber \\
 \one+P_{\rm e} & = 2^{n-1}(\ketbra{\psi^\text{odd}}{\psi^\text{odd}} + \ketbra{\tilde{\psi}^\text{odd}}{\tilde{\psi}^\text{odd}}).
\end{eqnarray}
From Observation 1 it follows that both $\one+P_{\rm e}$ and $P_{\rm o}$ are 
diagonal in the same basis and commute. The eigenvalues then can be read off. 
Furthermore, Eq.~(\ref{eq:result1}) can be directly verified in the common 
eigenbasis. 
\qed

The fact that $P_{\rm e}$ is given by $P_{\rm o}$ for pure states 
can be restated in the language of uniqueness: Pure qubit states of 
an odd number of parties are uniquely determined among pure states 
(UDP) by the odd correlations. %Later we will even show that they are
%determined among all mixed states (UDA) by  $P_{\rm o}$. 
This leads to 
the converse question of whether these states are also determined by the 
even correlations $P_{\rm e}$. The answer to this question is negative,
but the set of compatible states is rather small.

\noindent
{\bf Remark 3.}
{\it
Given the even correlations $P_{\rm e}$ of a pure $n$-qubit state 
$\ket{\psi^\text{odd}}$, the set of admissible odd correlations $P_{\rm o}$ to retrieve
a pure state again is a two-parameter family. The proof 
is given in Appendix A.}

%%%%%%%%%%%%%%%%%%%%%%%%%%%%%%%%%%%%%%%%%%%%%%%%%%%%%%%%%%%%%%%%%%%%
{\it Application I: Uniqueness among all states.---} 
So far, we have shown that for an odd number of parties, the odd correlations 
uniquely determine the state among all pure states. This is already a 
generalization of previous results \cite{diosi2004three, wyderka2017almost}, 
but one can ask the more general question, whether a state is determined by 
the correlations among all states (UDA), pure or mixed \cite{chen2013uniqueness}. For that question, 
some results are known \cite{linden2002almost, jones2005parts}, which we can
generalize now.

\noindent
{\bf  Corollary 4.}
{\it Consider a pure qubit state $\ket{\psi}$ of $n$ parties where $n$ is odd. 
Then the state is uniquely determined among all mixed states by $P_{\rm o}.$
}

{\it Proof.}
Recall that in the even-odd decomposition, the state reads
$
\ketbra{\psi}{\psi}=(\one+P_{\rm e}+P_{\rm o})/2^n.
$
Suppose there were a mixed state $\rho$ with the same odd correlations. Then 
we could write it as a convex sum of pure states, 
\begin{equation}
\rho=\sum_{i}p_{i}\frac{1}{2^{n}}(\one+P_{\rm e}^{(i)}+P_{\rm o}^{(i)}),
\end{equation}
where $\sum_{i}p_{i}=1$ and $\sum_{i}p_{i}P_{\rm o}^{(i)}=P_{\rm o}$.
From Observation~2 we know that $P_{\rm o}$ has two non-vanishing 
eigenvalues $\lambda_{o_{\pm}}=\pm2^{n-1}$, and the same holds for 
every $P_{\rm o}^{(i)}$ as they originate from pure states. Because 
the largest eigenvalue of the sum equals the sum of all the maximal
eigenvalues, all $P_{\rm o}^{i}$ must share the same corresponding 
eigenvector. The same is true for the second and lowest eigenvalue.
Thus, $P_{\rm o}^{(i)}=P_{\rm o}^{(j)}$  for all $i,j$ follows.
As the $P_{\rm e}^{(i)}$ are uniquely determined by the 
$P_{\rm o}^{(i)}$, they also coincide and therefore
$
\rho=\ketbra{\psi}{\psi}.
$
\qed

This result can be seen as a generalization of Ref.~\cite{linden2002almost},
where it was shown that almost all three-qubit states are determined among all states by $P_1$ and
$P_2$. Corollary 4 shows that all three-qubit states are determined among all states by $P_1$ and 
$P_3$, and it is remarkable that this generalizes to all odd numbers of 
parties.

An immediate consequence of Corollary 4 is that all pure states of an odd number of parties are
unique ground states of odd-body Hamiltonians. More precisely, choosing $H = -P_{\rm o} = 2^{n-1}(\ketbra{\tilde{\psi}^\text{odd}}{\tilde{\psi}^\text{odd}}
 - \ketbra{\psi^\text{odd}}{\psi^\text{odd}})$ yields a specific example of such a Hamiltonian.

{\it Results for an even number of qubits.---}
We now turn to the case of even $n$, and throughout this section, 
$\ket{\psi^\text{even}}$ denotes a pure state on an even number of 
qubits.
Although in this case $\ket{\psi}$ and $\ket{\tilde{\psi}}$ do not 
need to be perpendicular, one can gain some insight on the even and odd 
components of the correlations. We denote the overlap by 
$\vert \braket{\tilde{\psi}|\psi} \vert= \alpha$ with a positive number
$\alpha$ such that $\trace(\rho\tilde{\rho})=\alpha^{2}$. \kommentar{Note that 
the phase $\phi$ is not a physical property of the state, as changing 
the state $\ket{\psi}$ to $e^{i\beta}\ket{\psi}$ changes the value of 
$\phi$ while describing the same state. 
The value $\alpha$, however, is physical and the properties of the 
state depend on it.}

For pure states and $n=2$, $\alpha$ is 
just the concurrence \cite{wootters1998entanglement}. For pure states 
with $n\geq2$, $\alpha$ is known as the $n$-concurrence of a state and 
is known to be an entanglement monotone \cite{wong2001potential}. 
For our purpose, we need to distinguish three cases: The case where 
$\alpha=0$, the case of $0<\alpha<1$ and that of $\alpha=1$.

If $\alpha=0$, we recover the case of an odd number of qubits and 
the same results are valid. Examples for such states are the $W$-state,
$\ket{W}=(\ket{0\ldots01}+\ldots+\ket{10\ldots0})/{\sqrt{n}}$,
and all completely separable states. In this case, all the results
from the previous sections apply and $P_{\rm o}$ determines  
$P_{\rm e}$.

If $\alpha=1$, $\ket{\psi}\propto\ket{\tilde{\psi}}$, which means 
that there are only even correlations present in $\ket{\psi}$ and 
$P_{\rm o}=0$. In this case, the even correlations are not determined 
by the odd correlations at all. One prominent example for such a 
state is the $n$-party Greenberger-Horne-Zeilinger (GHZ) state, 
$\ket{\text{GHZ}}=(\ket{0\ldots0}+\ket{1\ldots1})/\sqrt{2}$.

If $0<\alpha<1$, even though the results from the previous chapter 
do not apply, the spectrum of $P_{\rm e}$ is still rather fixed, 
leading to the following:

\noindent
{\bf Observation 6.}
{\it Let $\ket{\psi^\text{even}}$ be a pure qubit state with
$|\braket{\psi^\text{even}|\tilde{\psi}^\text{even}}|^2=\alpha^2 \neq0$. Write $\ket{\psi^\text{even}}$ in
the even-odd decomposition as in Eq.~(\ref{evenodddecomp}). Then
\\
(1) the even correlations $P_{\rm e}$ uniquely determine the 
odd correlations $P_{\rm o}$ up to a sign; and
\\
(2) the family of pure states having the same odd correlations $P_{\rm o}$ as $\ket{\psi^\text{even}}$
is one-dimensional. The even correlations can be parametrized in terms of $P_{\rm o}$.}

The proof of this Observation is given in Appendix B. The results of all
the previous observations are summarized in Table~\ref{table-summary}.

\begin{table}[t]
\begin{center}
\begin{tabular}{|c||c|c|}
\hline
 & $n$ even and $0<\alpha<1$ & $n$ odd or $\alpha=0$\tabularnewline
\hline 
\hline 
$P_{\rm o}$ given & One-dimensional  & $P_{\rm e}$ is uniquely \tabularnewline
 & solution space for $P_{\rm e}$ & determined\tabularnewline
% & (Eq.~(\ref{eq:result4})) & (Eq.~(\ref{eq:result1}))\tabularnewline
\hline 
$P_{\rm e}$ given & $\pm P_{\rm o}$ is uniquely de- & Two-dimensional \tabularnewline
 & termined up to the sign & solution space for $P_{\rm o}$\tabularnewline
% & (Eq.~(\ref{eq:result3})) & (Eq.~(\ref{eq:result2}))\tabularnewline
\hline 
\end{tabular}
\par\end{center}
\caption{Summary of the relations between the even and odd components of 
pure state correlations as derived in Observation 2, Remark 3 and Observation 6. 
The detailed relations can be found in the corresponding proofs. Additionally, 
if $n$ is even and $\alpha=1$, the state exhibits only even correlations and given 
$P_{\rm e}$, only $P_{\rm o}=0$ is compatible.
}
\label{table-summary}
\end{table}

A statement similar to Corollary 4 is not true for an even number of 
parties with $\alpha\neq0$, as the family of mixed states 
$p\rho+(1-p)\tilde{\rho}=[\one+P_{\rm e}+(2p-1)P_{\rm o}]/2^n$ shares the same
even body correlations, unless $\alpha=1$, in which case $P_{\rm o} = 0$
and the state is determined.

As a final remark, note that pure states mixed with white noise
can be reconstructed as well from knowledge of $P_{\rm o}$ ($P_{\rm e}$)
for $n$ odd ($n$ even), as the noise parameter can be deduced from the eigenvalues
of the operators.

%%%%%%%%%%%%%%%%%%%%%%%%%%%%%%%%%%%%%%%%%%%%%%%%%%%%%%%%%%%%%%%

{\it Application II: Ground states.---} 
Some of our findings can be related to the Kramers theorem \cite{kramers1930theorie}.
Consider a Hamiltonian that contains even-body interactions
only, such as the Ising model without external field or the
$t$-$J$-model. A unique ground state of such a Hamiltonian must have
even correlations only.
This, however, is not possible if $n$ is odd, in which case
odd correlations must be present according to Eq.~(\ref{eq:result1}).
On the other hand, if $n$ is even, then the ground state must belong to the class of
even states, i.e.,~$\alpha=1$.
Second, consider Hamiltonians with odd-body interactions only. The
ground-state energy of such Hamiltonians is a function of $P_{\rm o}$
only. Thus, a unique ground state for $n$ even can only be a state
which is determined uniquely by $P_{\rm o}$, which are exactly the states
perpendicular to their inverted states, i.e.,~having $\alpha=0$ like the
$W$-state or product states.

%%%%%%%%%%%%%%%%%%%%%%%%%%%%%%%%%%%%%%%%%%%%%%%%%%%%%%%%%%%%%%%%%
{\it Application III: Unitary time evolution.---} 
Another application concerns the orbits of certain states under the time
evolution of Hamiltonians. Here, our approach allows one to re-derive and 
understand previous results from Ref.~\cite{bullock2005time}, where 
a completely different approach was used. Consider a Hamiltonian $H_{\rm o}$ 
consisting of odd-body interactions only. Then, any operator $P$ evolves 
in time as
\begin{equation}
P(t)=e^{-iH_{\rm o}t}Pe^{iH_{\rm o}t}
=\sum_{m=0}^{\infty}\frac{(-it)^{m}}{m!}[H_{o\rm },P]_{m},
\label{eq:comm}
\end{equation}
where $[H_{\rm o},P]_{m}:=[H_{\rm o},[H_{\rm o},P]_{m-1}]$ is the $m$-times
nested commutator with $[H_{\rm o},P]_{0}=P$.

Now, recall that we denote by $\weight(T)$ the weight of a tensor product
of Pauli matrices. For these weights, Lemma 1 from 
Ref.~\cite{huber2017absolutely}, adapted for the case of 
commutators, can be used. It states that for the weight of the 
commutator of two tensor products $S$ and $T$ one has that:
\begin{equation}
\weight([S,T])\equiv\weight(S)+\weight(T)+1\quad(\text{mod }2),
\end{equation}
provided that the commutator does not vanish. This lemma
encodes the commutator rules of the Pauli matrices.
Therefore, by linearity, commuting two odd or two even Hermitian operators yields an odd 
operator, while commuting an even and an odd operator yields an even operator.

Consider, for example, the three-qubit operators 
$S=\sigma_x\otimes\sigma_y\otimes \sigma_z + \one \otimes \one \otimes \sigma_y$ 
and $T=\one \otimes \sigma_x \otimes \sigma_z$. Then, $S$ has odd and $T$ has 
even weight. Their commutator is given by
$[S,T]=-2i\sigma_x\otimes \sigma_z \otimes \one + 
2i \one \otimes \sigma_x \otimes \sigma_x$, which has even weight.

Thus, 
if $H$ and $P$ are odd, all the nested commutators in Eq.~(\ref{eq:comm}) are odd too, 
and $P(t)$ stays odd for all times $t$.
On the other hand, if $H$ is odd but $P$ is even, then $P(t)$ remains even. 
By Eqs.~(\ref{evenodddecomp}) and (\ref{evenodddecomp2}), $\tilde\rho$ evolves too as
$\exp(-iHt)\tilde\rho\exp(iHt)$, as the state inversion and the time evolution
commute in this case.
Therefore, given a quantum state $\rho$, the overlap
$\alpha^{2}=\trace(\rho\tilde{\rho})$ stays constant for all times. 
This is also true for mixed states. In that case, the result also holds for the
$n$-concurrence $C_n$, given by the convex roof construction
for $\alpha$, as the value of $\trace(\rho\tilde{\rho})$ stays 
constant for any decomposition of $\rho$ into a sum over 
pure states \cite{uhlmann2000fidelity}. So we have the following. 

\noindent
{\bf Observation 7.}
{\it Any quantum state $\rho(t)$, whose time evolution
is governed by an odd-body interacting Hamiltonian has a constant
value of $\alpha$ and $C_n$.}

This result can be useful as follows: Recent experiments enabled 
the  observation of the spreading of quantum correlations under interacting 
Hamiltonians for systems out of thermal equilibrium \cite{cheneau, blattspreading}. 
Observation 7 shows that large classes of Hamiltonians preserve certain properties 
of a quantum state and deviations thereof may be used to characterize the 
actually realized Hamiltonian. For instance, Refs.~\cite{pachos, buechler} 
proposed methods to engineer Hamiltonians with three-qubit interactions only. 
Experimentally, the $n$-concurrence is not easy to measure; however, bounds
can be found with simple methods \cite{schmid2008experimental, vanenk2009direct, zhang2016evaluation}.
A simple scheme that detects even-body terms in the Hamiltonian is the following.

Start with any state $\ket{\psi(0)}$ with zero $n$-concurrence and let it evolve under the Hamiltonian in question.
After a fixed time $t_0$, the state can be decomposed as
\begin{equation}
\ket{\psi(t_0)} = \sqrt{F}\ket{\text{GHZ}} + \sqrt{1-F}\ket{\chi}
\end{equation}
with $\braket{\text{GHZ}|\chi} = 0$. The $n$-concurrence of the state is given by
\begin{eqnarray}
C_n(\ket{\psi(t_0)}) & = & \vert \braket{\psi(t_0)|\tilde{\psi}(t_0)} \vert \nonumber \\
                     & = & \vert F \braket{\text{GHZ}|\text{GHZ}} + (1-F)\braket{\chi|\sigma_y^{\otimes n}|\chi^*}    \nonumber \\
                     &   & + \sqrt{F(1-F)}(\braket{\text{GHZ}|\sigma_y^{\otimes n}|\chi^*} + \text{H.c.})\vert \nonumber \\
                     & = & \vert F + (1-F)\braket{\chi|\tilde{\chi}} \vert,
\end{eqnarray}
as $\braket{\text{GHZ}|\sigma_y^{\otimes n}|\chi^*} = \braket{\text{GHZ}|\chi^*} = \braket{\text{GHZ}|\chi}^* = 0$.
The right hand side is always lower bounded by
\begin{equation}
C_n(\ket{\psi(t)}) \geq F - (1-F).
\end{equation}
If $F > 50\%$, the concurrence is non-zero and even-body interactions must have been present.
Therefore, low-concurrence states cannot approximate the GHZ state under the time evolution with odd-body Hamiltonians.

Combining this result with the one about ground states of odd-body Hamiltonians,
we arrive at the following

\noindent
{\bf Observation 8.}
{\it If $n$ is even, it is not possible to produce a GHZ state from a $W$-state
(or any state with $C_n=0$) by unitary or adiabatic time evolution under
Hamiltonians with odd interactions only.}

{\it Application IV: Entanglement detection.---}
%As a final application we consider the problem of entanglement detection. 
The results of this paper yield insight into the structure of pure quantum
states that is still subject to ongoing research \cite{goyeneche2015five}.

Consider a pure state of $n$ qubits with $n$ being odd. Suppose that the 
odd correlations $P_{1},P_{3},\ldots,P_{n-2}$ are given. If the state is 
biseparable, there are ${(n-1)}/{2}$ different possibilities of biseparation: 
It could be biseparable along a cut between one qubit and the other $n-1$ 
qubits, or between two qubits and $n-2$, etc., up to ${(n-1)}/{2}$ qubits 
and ${(n+1)}/{2}$ qubits.
The first case can be tested for by checking for each party whether
the corresponding one-particle reduced state is pure. This can be
done due to knowledge of $P_{1}$. The second case, namely, two qubits
vs.~$n-2$ qubits can be tested by assuming that the $(n-2)$-qubit
state is pure and trying to reconstruct the appropriate even correlations.
According to Corollary 4, this is only possible if the
state was indeed pure. This procedure can be applied for all other
splittings as well. 
Thus, the information on genuine multipartite entanglement in pure
states is embodied in the odd correlations $P_{1},P_{3},\ldots,P_{n-2}$
already, where no knowledge of the highest correlations $P_{n}$ is
needed. This is in contrast to the case of mixed states.

%%%%%%%%%%%%%%%%%%%%%%%%%%%%%%%%%%%%%%%%%%%%%%%%%%%%%%%%%%%%%%55

{\it Possible generalizations.---}

While the results obtained in this paper are valid for qubit systems
only, some extensions to higher-dimensional systems are possible,
as we will discuss now.
As stated in the main text, the state inversion can be generalized to systems of internal
dimension $d$, as discussed in \cite{eltschka2017distribution},
\begin{equation}
\tilde{\rho}:=\one^{\otimes n}-\sum_{i=1}^{n}\rho_{(i)}+\sum_{i<j}\rho_{(ij)}-\ldots\pm\rho.\label{eq:rhotilde-1}
\end{equation}
This yields a positive operator \cite{hall2005multipartite, rungta2001universal}, which can be normalized to a proper state. In the
Bloch decomposition, the inversion reads
\begin{eqnarray}
\tilde{\rho} & = & \frac{(d-1)^n}{d^{n}}\sum_{j=0}^{n}\left(\frac{1}{1-d}\right)^{j}P_{j}.\label{eq:rhotildebloch-1}
\end{eqnarray}
However, for $n>1$ and $d>2$ pure states do not stay pure under the state inversion.
Thus, the state inversion cannot be represented
as an operator acting on vectors in Hilbert space anymore, but only
as a channel.
Nevertheless, this generalization has recently been used to
gain insight into the distribution of entanglement in higher
dimensional many-body systems \cite{eltschka2017distribution}.

Another generalization concerns the nature of the inversion operator. Instead
of flipping the sign of all non-trivial Pauli operators, one can generalize
this to only flipping certain ones. The most general
form of such an operator acting on a single qubit reads
\begin{equation}
F_{\alpha}=iC(i\alpha_{0}\one+\sum_{i=1}^{3}\alpha_{i}\sigma_{i})
\end{equation}
where the four dimensional vector $\vec{\alpha}$ is normalized. The
choice $\alpha=(0,0,1,0)^{\text{T}}$ corresponds to the flip considered
above (the signs of all Pauli matrices are flipped), whereas $\alpha=(1,0,0,0)^{\text{T}}$
flips just $\sigma_{y}$ (which corresponds to a transposition of the state), $\alpha=(0,1,0,0)^{\text{T}}$ flips $\sigma_{z}$,
and $\alpha=(0,0,0,1)^{\text{T}}$ flips $\sigma_{x}$. Other values
of $\vec{\alpha}$ correspond to superpositions of these flips. Indeed,
$F_{\alpha}\ket{\psi}$ is a pure state again. For example, setting
$\alpha=(0,0,0,1)^{\text{T}}$ allows for a decomposition of states
by the number of $\sigma_{x}$ appearing in each term, thus, $P_{\rm e}$
would consist of all terms with an even number of $\sigma_{x}$. Using
this decomposition, analogous results can be derived with similar
uniqueness properties.

%%%%%%%%%%%%%%%%%%%%%%%%%%%%%%%%%%%%%%%%%%%%%%%%%%%%%%%%%%%%%%55

{\it Discussion.---}
We introduced the decomposition of multipartite qubit states in terms of even 
and odd correlations. For pure states, we showed that the even and odd correlations 
are deeply connected, and often one type of correlations determines
the other. This allowed us to prove several applications, ranging from the 
unique determination of a state by its odd correlations to invariants under
Hamiltonian time evolution and entanglement detection. 

For future work, it would be highly desirable to generalize the approach
to higher-dimensional systems. Some facts about state inversion are 
collected in the previous section, but developing a general theory seems challenging. 
Furthermore, it may be very useful if one can extend our theory  to a 
quantitative theory, where the correlations within some subset of particles
are measured with some correlation measure and then monogamy relations between 
the  different types of correlations are developed. 

{\it Note added.---}
In a previous version of this manuscript,
we claimed that 
``[...] if a state $\ket{\psi}$ is uniquely determined among all states by certain sets of correlations (for example, odd-body correlations), 
then $\ket{\psi}$ is the unique ground state of some Hamiltonian having interaction terms from that set only.''
This statement is not correct, as a recently found analytical counterexample on six qubits shows~\cite{karuvadeUnique}.

The error in the reasoning occured after Lemma 5: 
The conclusion ``As $\mathcal{R}(\ket{\psi}\!\bra{\psi})$ is extremal, there exists a linear witness $L$ in the projected space $\mathcal{R}(\mathcal{O})$ of all Hermitian operators $\mathcal{O}$ with $\braket{\psi|L|\psi}$ being minimal'' is wrong. An explanation of the fallacy is already given in Ref.~\cite{chenFrom} (cf. Fig.~1): an explicit two-dimensional set is constructed, 
in which some extremal points can  not be separated from all other extremal points by any linear witness. 
Thus, any such a witness cannot detect a single state uniquely. 
In our context, such a witness cannot be used as a Hamiltonian
with non-degenerate ground state space.

We removed the wrong statement and the now unnecessary Lemma~5, and inserted a correct argument for the claim that 
``[...] all pure states of an odd number of parties are unique ground states of odd-body Hamiltonians.''

\begin{acknowledgments}
{\it Acknowledgments.---}
We thank Jens Siewert, Gavin Brennen and Lorenza Viola for fruitful discussions. 
This work was supported
by the Swiss National Science Foundation (Doctoral Mobility Grant No.
165024), the DFG, the ERC (Consolidator Grant No. 683107/TempoQ),
and the House of Young Talents Siegen.
\end{acknowledgments}
%%%%%%%%%%%%%%%%%%%%%%%%%%%%%%%%%%%%%%%%%%%%%%%%%%%%%%%%%%%%%%%%5

\appendix

\section*{Appendix}

\subsection*{A: Proof of Remark 3}

\noindent
{\bf Remark 3.}
{\it
Given the even correlations $P_{\rm e}$ of a pure $n$-qubit state 
$\ket{\psi^\text{odd}}$, the set of admissible odd correlations $P_{\rm o}$ to retrieve
a pure state again is a two-parameter family.}

{\it Proof.}
Let $\rho=\ketbra{\psi^\text{odd}}{\psi^\text{odd}}$ and $\tilde{\rho}=\ketbra{\tilde{\psi}^\text{odd}}{\tilde{\psi}^\text{odd}}$,
and write $\one+P_{\rm e}=2^{n-1}(\rho+\tilde{\rho})$. Thus, the
eigenvectors with eigenvalue $2^{n-1}$ of $\one+P_{\rm e}$ are a superposition
of $\ket{\psi^\text{odd}}$ and $\ket{\tilde{\psi}^\text{odd}}$. Given 
only $P_{\rm e}$, one can choose any of its eigenvectors $\ket{\eta}$ from 
the two-dimensional subspace of eigenvalue $2^{n-1}-1$. As $\ket{\tilde{\eta}}$
is orthogonal to $\ket{\eta}$, it follows that
$\one+P_{\rm e} = 2^{n-1}(\ketbra{\eta}{\eta} + \ketbra{\tilde\eta}{\tilde\eta})$.
Therefore, every choice of an eigenvector gives rise to compatible correlations
$P_{\rm o}^{(r)}$ via \begin{eqnarray}
P_{\rm o}^{(r)} & = & 2^{n-1}(\ketbra{\eta}{\eta}-\ketbra{\tilde{\eta}}{\tilde{\eta}}),
\end{eqnarray}
resulting in the total state $\rho=\ketbra{\eta}{\eta}.$ By fixing 
one of the eigenstates $\ket{\eta}$, one can parametrize all valid 
solutions by 
\begin{eqnarray}
P_{\rm o}^{(r)}(\theta, \phi) & = & 2^{n-1}[\cos\theta(\ketbra{\eta}{\eta}-\ketbra{\tilde{\eta}}{\tilde{\eta}})\nonumber \\
 &  & \phantom{2^{n-1}}\!\!\!\!+\sin\theta(e^{i\phi}\ketbra{\tilde{\eta}}{\eta}+e^{-i\phi}\ketbra{\eta}{\tilde{\eta}})]\label{eq:result2}
\end{eqnarray}
for all real valued $\theta$ and $\phi$.
\qed

%%%%%%%%%%%%%%%%%%%%%%%%%%%%%%%%%%%%%%%%%%%%%%%%%%%%%%%%%%%%%%%%%%55
\subsection*{B: Proof of Observation 6}
\noindent
{\bf Observation 6.}
{\it Let $\ket{\psi^\text{even}}$ be a pure qubit state with
$|\braket{\psi|\tilde{\psi}}|^2=\alpha^2 \neq0$. Write $\ket{\psi^\text{even}}$ in
the even-odd decomposition as in Eq.~(\ref{evenodddecomp}). Then
\\
(1) the even correlations $P_{\rm e}$ uniquely determine the 
odd correlations $P_{\rm o}$ up to a sign; and
\\
(2) the family of pure states having the same odd correlations $P_{\rm o}$ as $\ket{\psi^\text{even}}$
is one-dimensional. The even correlations can be parametrized in terms of $P_{\rm o}$.}

{\it Proof.}
Let $\rho=\ketbra{\psi^\text{even}}{\psi^\text{even}}$.
Before proving the statements, we investigate the eigenvectors
and eigenvalues of $P_{\rm e}$ and $P_{\rm o}$.
As $\one+P_{\rm e}=2^{n-1}(\rho+\tilde{\rho})$, it
must be of rank two if $\alpha\neq1$. Thus, it has two non-vanishing eigenvalues,
lying in the span of $\ket{\psi}$ and $\ket{\tilde{\psi}}$. Calculating
\begin{eqnarray}
(\one+P_{\rm e})\ket{\psi} & = & 2^{n-1}(\ket{\psi}+\alpha e^{i\phi}\ket{\tilde{\psi}}),
\nonumber
\\
(\one+P_{\rm e})\ket{\tilde{\psi}} & = & 2^{n-1}(\ket{\tilde{\psi}}+\alpha
e^{-i\phi}\ket{\psi})
\end{eqnarray}
yields the two non-vanishing eigenvalues
\begin{equation}
1+\lambda_{{\rm e}_{\pm}}=2^{n-1}(1\pm \alpha)
\end{equation}
and the corresponding orthonormal eigenvectors
\begin{equation}
\ket{e_{\pm}}=\frac{1}{\sqrt{2(1\pm \alpha)}}
(\ket{\psi}\pm e^{i\phi}\ket{\tilde{\psi}}).
\end{equation}
We can also determine the action of $P_{\rm o}$ on these eigenvectors,
which reveals that it is purely off-diagonal in the eigenbasis of $P_{\rm e}$,
\begin{equation}
P_{\rm o}\ket{e_{\pm}}  =  2^{n-1}(\rho-\tilde{\rho})\ket{e_{\pm}}
  =  2^{n-1}\sqrt{1-\alpha^{2}}\ket{e_{\mp}}.
\end{equation}
Thus, the eigenvectors of $P_{\rm o}$ are given by
\begin{equation}
\ket{o_{\pm}}=\frac{1}{\sqrt{2}}(\ket{e_{+}}\pm\ket{e_{-}})\label{eq:ket_opm}
\end{equation}
and the eigenvalues are given by
\begin{equation}
\lambda_{{\rm o}_{\pm}}=\pm2^{n-1}\sqrt{1-\alpha^{2}}.\label{eq:lambda_o_n_even}
\end{equation}
We are now in position to prove the claims.
Let us prove statement two first:

(2) By assumption, $P_{\rm o}$ is known. The eigenvalues determine
the overlap $\alpha$ by Eq.~(\ref{eq:lambda_o_n_even}).
Knowledge of $\alpha$ fixes the eigenvalues of any admissible reconstructed
$P_{\rm e}^{(r)}$. The admissible eigenvectors of $P_{\rm e}^{(r)}$ can be 
obtained from Eq.~(\ref{eq:ket_opm}) to read 
\begin{equation}
\ket{e_{\pm}}=\frac{1}{\sqrt{2}}(\ket{o_{+}}\pm\ket{o_{-}})\text{.}
\end{equation}
However, the eigenvectors $\ket{o_{\pm}}$ are only unique
up to a phase. Taking into account this extra phase while omitting
a global phase yields
\begin{equation}
\ket{e_{\pm}}=\frac{1}{\sqrt{2}}(\ket{o_{+}}\pm e^{i\varphi}\ket{o_{-}}).
\end{equation}
This allows us to write all compatible even correlations as 
\begin{eqnarray}
\one+P_{\rm e}^{(r)} & = & (1+\lambda_{{\rm e}_{+}})\ketbra{e_{+}}{e_{+}}+(1+\lambda_{{\rm e}_{-}})\ketbra{e_{-}}{e_{-}}\nonumber \\
 & = & 2^{n-1}(\ketbra{o_{+}}{o_{+}}+\alpha e^{-i\varphi}\ketbra{o_{+}}{o_{-}}\nonumber \\
 &  & \phantom{2^{n-1}}\!\!\!\!\!+\ketbra{o_{-}}{o_{-}}+\,\,\,\,\alpha e^{i\varphi}\ketbra{o_{-}}{o_{+}})\,.\label{eq:result4}
\end{eqnarray}
This is a one-dimensional space of admissible
reconstructed even correlations, parametrized by $\varphi$.

We now show the first statement:

(1) Assume that now $P_{\rm e}$ is
given. Can we uniquely reconstruct in the odd correlations $P_{\rm o}$
from knowledge of $P_{\rm e}$? Unfortunately, the eigenvectors $\ket{e_{\pm}}$
are again only determined up to a phase. Therefore, every reconstructed
operator $P_o^{(r)}$ of the form
\begin{eqnarray}
P_{\rm o}^{(r)} & = & \lambda_{o_{+}}\ketbra{o_{+}}{o_{+}}+\lambda_{-}\ketbra{o_{-}}{o_{-}}\\
 & = & \lambda_{o_{+}}(e^{i\varphi}\ketbra{e_{+}}{e_{-}}+e^{-i\varphi}\ketbra{e_{-}}{e_{+}})\label{eq:result3}
\end{eqnarray}
for all $\varphi\in\mathbb{R}$ would be a valid operator, such that
\begin{equation}
\frac{1}{2^{n}}(\one+P_{\rm e}+P_{\rm o}^{r})
\end{equation}
is a pure state again. However, only certain choices of $\varphi$
recreate a $P_{\rm o}$ which exhibits solely odd correlation in Bloch
decomposition. This can be seen as follows: As shown in Lemma 9 below,
$\ketbra{e_{\pm}}{e_{\pm}}$ can only exhibit even correlations.
This means that $\ket{e_{\pm}}$ are eigenvectors of the inversion
operator $F$ introduced above, i.e.~$F\ket{e_{\pm}}\propto\ket{e_{\pm}}$.
Recall that for $n$ even, $F^{\dagger}=F$. Thus,
$F\ketbra{e_{+}}{e_{-}}F^{\dagger}=e^{i\Lambda}\ketbra{e_{+}}{e_{-}}$
for some $\Lambda$. The condition that $P_{\rm o}^{(r)}$ contains only
odd correlations can be written as
\begin{equation}
P_{\rm o}^{(r)}+\tilde{P}_{\rm o}^{(r)}=P_{\rm o}^{(r)}+FP_{\rm o}^{(r)}F^{\dagger}=0.
\end{equation}
Eq.~(\ref{eq:result3}) translates this to
\begin{equation}
e^{i\varphi}+e^{-i(\varphi-\Lambda)}=0,
\end{equation}
which exhibits exactly two solutions for $\varphi$. Thus, there are
only two possible reconstructions $P_{\rm o}^{(r)}$, corresponding to the
original $P_{\rm o}$ and its negation, $-P_{\rm o}$. 
\qed

All that is left is to show the used assumption that the eigenvectors
$\ket{e_{\pm}}$ exhibit only even correlations. Note, that this is
a special case of Kramers theorem \cite{kramers1930theorie}, stating that
the eigenstates of a Hamiltonian exhibiting even correlations only is
either at least two-fold degenerate or exhibits itself only even correlations.

\noindent
{\bf Lemma 9.}
{\it
Let $P=\lambda_{+}\ketbra{p_{+}}{p_{+}}+\lambda_{-}\ketbra{p_{-}}{p_{-}}$
be a Hermitian operator which exhibits only even correlations in the Bloch
decomposition, $\braket{p_{+}|p_{-}}=0$ and $\lambda_{-}<\lambda_{+}$.
Then $\ketbra{p_{+}}{p_{+}}$ and $\ketbra{p_{-}}{p_{-}}$ also exhibit
only even correlations.}

{\it Proof.}
We regard $P$ as a Hamiltonian with the unique ground state $\ket{p_-}$.
As $P$ has even correlations only, $FPF^{\dagger}=P$. Thus
\begin{eqnarray}
\lambda_{-} & = & \trace(P\ketbra{p_{-}}{p_{-}})=\trace(FPF^{\dagger}\ketbra{p_{-}}{p_{-}})\nonumber \\
 & = & \trace(PF\ketbra{p_{-}}{p_{-}}F^{\dagger}),
\end{eqnarray}
as $F^{\dagger}=F$ if $n$ is even. Thus, also $F\ket{p_{-}}$ is a ground state
of $P$. As by assumption the ground state is unique, $F\ket{p_{-}}\propto\ket{p_{-}}$
must hold true and therefore, $\ketbra{p_{-}}{p_{-}}$ exhibits only
even correlations. This implies that also $\ketbra{p_{+}}{p_{+}}$
has even correlations only.
\qed

\end{document}